\documentclass[manuscript]{aastex}

\usepackage{graphicx}
\usepackage{amssymb}
\usepackage{natbib}
\usepackage{amsmath}
\usepackage{url}

\slugcomment{}

\shorttitle{Multiwavelength imaging and spectroscopy of chromospheric evaporation in an M-class solar flare}
\shortauthors{Veronig et al.}

\begin{document}

\title{Multiwavelength imaging and spectroscopy of chromospheric evaporation in an M-class solar flare}

\author{A.M. Veronig}
\affil{Institute of Physics, University of Graz,
    Universit\"atsplatz 5, A-8010 Graz, Austria}
    \email{asv@igam.uni-graz.at}

\author{J. Ryb\'ak}
\affil{Astronomical Institute, Slovak Academy of Sciences, SK-05960 Tatransk\'a Lomnica, Slovakia}

\author{P. G\"om\"ory}
\affil{Institute of Physics/Kanzelh\"ohe Observatory, University of Graz,
     A-9521 Treffen, Austria;
     Astronomical Institute, Slovak Academy of Sciences, SK-05960 Tatransk\'a Lomnica, Slovakia}

\author{S. Berkebile-Stoiser}
\affil{Institute of Physics, University of Graz,
    Universit\"atsplatz 5, A-8010 Graz, Austria}
    
\author{M. Temmer}
\affil{Institute of Physics, University of Graz,
    Universit\"atsplatz 5, A-8010 Graz, Austria;
    Space Research Institute, Austrian Academy of Sciences, Schmiedlstra{\ss}e 6, A-8042 Graz, Austria}
    
\author{W. Otruba}
\affil{Institute of Physics/Kanzelh\"ohe Observatory, University of Graz,
    A-9521 Treffen, Austria}
     
\author{B. Vr\v{s}nak}
\affil{Hvar Observatory, Faculty of Geodesy, Ka\v{c}i\'ceva 26, 1000 Zagreb, Croatia}
    
\author{W. P\"otzi}
\affil{Institute of Physics/Kanzelh\"ohe Observatory, University of Graz,
    A-9521 Treffen, Austria}

\author{D. Baumgartner}
\affil{Institute of Physics/Kanzelh\"ohe Observatory, University of Graz,
    A-9521 Treffen, Austria}

\begin{abstract}
We study spectroscopic observations of chromospheric evaporation mass flows in comparison to the energy input by electron beams 
derived from hard X-ray data for the white-light M2.5 flare of 2006 July 6. The event was captured 
in high cadence spectroscopic observing mode by SOHO/CDS combined with high-cadence imaging at various wavelengths in the visible, EUV and X-ray domain during 
the joint observing campaign JOP171. During the flare peak, we observe downflows in the He\,{\sc i} and O\,{\sc v} lines 
formed in the chromosphere and transition region, respectively, and simultaneous upflows in the hot coronal 
Si~{\sc xii} line. The energy deposition rate by electron beams derived from RHESSI hard X-ray observations is suggestive of explosive chromospheric evaporation, consistent
with the observed plasma motions. However, for a later distinct X-ray burst, where the site of the strongest energy deposition is exactly located 
on the CDS slit, the situation is intriguing. The O\,{\sc v} transition region line spectra show the evolution of double components, indicative of 
the superposition of a stationary plasma volume and upflowing plasma elements with high velocities (up to 280~km~s$^{-1}$) in single CDS pixels
on the flare ribbon. However, the energy input by electrons during this period is too small to 
drive explosive chromospheric evaporation. These unexpected findings indicate that the flaring transition region is much more 
dynamic, complex, and fine-structured than is captured in single-loop hydrodynamic simulations.
 
\end{abstract}



\section{Introduction}

Solar flares are thought to be the result of magnetic reconnection in
the corona, which releases vast amounts of energy within a few minutes 
(up to hours in the largest events). 
A substantial fraction of the released energy goes initially into the acceleration of suprathermal particles.  
Nonthermal electrons streaming down along the loops from a coronal acceleration region 
are stopped in the lower atmosphere due to the steep density increase in the transition 
region toward the chromosphere. There they deposit the bulk of their kinetic energy in Coulomb collisions
with the ambient thermal electron population \citep{1971SoPh...18..489B,1976SoPh...50..153L}. 
As a consequence of this energy input, the chromospheric  
plasma is strongly heated and expands upward, filling the coronal loops, which are then 
observed via their enhanced soft X-ray (SXR) and extreme ultraviolet (EUV) radiation. 
This process, known as ``chromospheric evaporation'', was first proposed by \cite{1968ApJ...153L..59N} 
to explain the observed correlation between the thermal and cumulated nonthermal flare emission 
\cite[``Neupert effect''; see e.g.][]{1993SoPh..146..177D,2002A&A...392..699V}. 

Strongly blue-shifted components and line-asymmetries during the flare impulsive 
phase evidencing plasma upflows with speeds
up to $\sim$300--400~km~s$^{-1}$ have been first observed in spatially unresolved SXR spectroscopy 
from highly ionized ions formed at temperatures of $10^7$~K with
the Bent and Bragg Crystal Spectrometer (BCS) onboard Solar Maximum Mission \citep[SMM; e.g.][]{1980ApJ...239..725D,1982SoPh...78..107A,1988ApJ...324..582Z,1990ApJS...73..117D}. 
Thereafter, flare-induced plasma flows were observed in SXR line spectra with Yohkoh/BCS \citep[e.g.][]{1993AdSpR..13..303C,1993ApJ...419..418M,1994ApJ...421L..55B,1994ApJ...424..459W},
and in spatially resolved EUV line spectra by SOHO/CDS 
\cite[Coronal Diagnostics Spectrometer; e.g.][]
{2003ApJ...586.1417B,2003ApJ...588..596T,2004ApJ...613..580B,2005ApJ...625.1027K,zanna2006,milligan2006}. 
Hydrodynamic simulations of the atmospheric response to electron-beam heating confirmed that strong blue-shifts in hot
flare lines should be observed 
\cite[e.g.][]{1984ApJ...279..896N,macneice1984,fisher1985a,fisher1985b,fisher1985c,2005ApJ...630..573A,liu09}.
These models also predict a relation between the temperature and the expansion velocity of the
evaporated plasma, which was recently confirmed from observations of a C-class flare with the 
EUV Imaging Spectrometer (EIS) onboard Hinode covering a multitude of emission lines
in the temperature range 0.05--16~MK \citep{milligan09}.

It is important to note that these one-dimensional single-loop models predict 
that in the early flare phase the total line profile should be blue-shifted by several
hundred kilometers per second, but the majority of the observations revealed only a blue asymmetry of the 
spectral lines indicating the presence of a strong static component
\citep[e.g.,][and references therein]{2005ApJ...629.1150D}. 
However, the observations can be reconciled with the numerical
simulations if a multi-thread fine structure of the flaring atmosphere is accounted for
in the modeling \citep{1997ApJ...489..426H,2005ApJ...618L.157W,2005ApJ...629.1150D}. 
In the multi-loop scenario, chromospheric evaporation occurs sequentially 
in a number of individual threads that together make up a larger loop system that may appear 
as a single loop in X-ray or EUV images. As a consequence, the integrated emission 
of the newly heated loops (subject to chromospheric evaporation plasma flows) and the
previously heated loops (static) do not reveal bulk motions.

Important observational diagnostics of the energy input by electron beams is provided by
their hard X-ray (HXR) radiation, assumed to be thick-target bremsstrahlung of electrons 
scattered off the ions when impinging on the chromosphere \citep{1971SoPh...18..489B}. 
The bulk of the kinetic energy of the fast electrons is lost in Coulomb collisions with the ambient thermal electrons, which is efficiently heating the 
flaring chromosphere. Only a tiny fraction ($\sim$$10^{-5}$) of the kinetic energy of the electrons is converted to HXR radiation when they are decelerated  
in the fields of the ions. However, it is the spatial and spectral distribution of this HXR bremsstrahlung, which provides us with important 
insight into the acceleration, transport, and energetics of electron beams in solar flares.

Hydrodynamic simulations predict that depending on the incident energy flux, 
the atmosphere responds by one of two means \citep[e.g.][]{fisher1985a,fisher1985b,fisher1985c,mariska1989,1999ApJ...521..906A}.
If the energy deposition rate is too small to raise the chromospheric temperature beyond the 
peak of the radiative loss function at around
$10^5$~K, the pressure can be raised by no more than a factor of 10. 
In this case, the heated atmosphere moves slowly upwards at several tens 
of kilometers per second to adjust to a new equilibrium position,  so-called ``gentle chromospheric evaporation''.
Gentle evaporation may also be due to thermal conduction from the hot flaring corona, and
was observed in small flares as well as in the preflare and late phase of larger events \citep{1999ApJ...521L..75C,2004ApJ...613..580B,2006ApJ...642L.169M}.
For high energy flux densities ($\gtrsim$$10^{10}$
ergs~cm$^{-2}$~s$^{-1}$), the chromosphere is unable to radiate away the deposited energy at
a sufficient rate and is rapidly heated to coronal temperatures. In this case, the pressure increases by at 
least a factor of $10^2$ (up to $10^3$). Consequently, the chromospheric plasma 
expands explosively upward into the loop at velocities of several hundred kilometers per second, in a process called
``explosive chromospheric evaporation''. The overpressure of the hot evaporating plasma relative to the
underlying chromosphere pushes cooler, more dense material downward at velocities of several 
tens of kilometers per second, establishing momentum balance 
with the hot plasma upflows \citep[e.g.][]{fisher1985c}. Such momentum balance was deduced from 
some flare observations to within an order of magnitude \citep{1988ApJ...324..582Z,canfield90,teriaca2006}. 
In a recent paper, \cite{brosius09} studied the plasma flow behavior at different temperatures with CDS high-cadence spectroscopy
and finds a change from explosive to gentle evaporation during the M1.5 flare in his investigation. 
This change from explosive 
to gentle evaporation could be due to the atmosphere being heated and filled with evaporated material during the earlier 
event. This would reduce the amount of electron beam flux reaching the chromosphere during the second event, 
since due to the increased column density a larger part of low-energy electrons is already stopped in the coronal part of the 
loop \citep{brosius09}.

In this paper, we study a well observed M2.5 flare that occurred on 2006 July 6, and concentrate on the flare-induced  
chromospheric evaporation mass flows observed by CDS, in comparison to the energy input by electron beams 
derived from hard X-ray data and the results of hydrodynamic simulations. The flare was captured with
high-cadence spectroscopy across the southern flare ribbon in UV and EUV emission lines by SOHO/CDS 
in combination with multi-wavelength high-cadence imaging of the photosphere (TRACE white light, SOHO/MDI white light and magnetograms), 
the chromosphere (TRACE UV, Kanzelh\"ohe and Hvar H$\alpha$) and transition region and corona (TRACE EUV).
In addition, we also have RHESSI X-ray observations during the flare impulsive phase, which provide us with 
imaging and spectroscopy of the hot flaring plasma in the corona ($\gtrsim$$10^7$~K) and allow us to deduce the
energy input by flare-accelerated electron beams. 

The paper is structured as follows. In Sect.~2, we give a detailed description of the various data and instruments involved in 
the study, and describe the co-alignment of the different data sets. In Sect.~3 we present an event overview and our results structured into 
multi-wavelength imaging, CDS spectroscopy and RHESSI imaging and spectroscopy. In Sect. 4, we discuss our findings and how 
they relate to hydrodynamic simulations of the flaring atmosphere. Finally, we present our conclusions in Sect.~5.

\section{Data and data reduction}

The flare observations analysed in this paper were acquired during a coordinated
observing campaign performed during 2006 June 28 to July 12.
The SOHO Joint Observing Program JOP171 included the operation of 
the Transition Region and Coronal Explorer \citep[TRACE;][]{handy1999}, the 
Coronal Diagnostics Spectrometer \cite[CDS;][]{harrison1995} and the Michelson Doppler Imager (MDI;
\citeauthor{scherrer1995} \citeyear{scherrer1995}) onboard the Solar and Heliospheric Observatory \citep[SOHO;][]{1995SoPh..162....1D}, as well as the ground-based Dutch Open Telescope on La Palma
(DOT; \citeauthor{hammerschlag1998} \citeyear{hammerschlag1998}), the Hvar Observatory  
of the University of Zagreb (Croatia) and the Kanzelh\"ohe Observatory of the University of 
Graz (Austria).\footnote{Further 
details on the campaign can be found at our JOP171 web page: {http://www.astro.sk/\~{}choc/open/06\_dot/06\_dot.html} }
On July 6, an M2.5 flare occurred at a heliographic position of (S09$^\circ$, W34$^\circ$) in the target field of view of AR~10898, 
which was observed by all campaign instruments except DOT (due to bad weather conditions). 
For our study, we also use the hard X-ray observations of the Ramaty High Energy Solar Spectroscopic Imager (RHESSI; \citeauthor{lin2002} \citeyear{lin2002}).

\subsection{SOHO CDS, MDI and EIT}

During JOP171, the Normal Incidence Spectrometer (NIS) on CDS \citep{harrison1995}
provided simultaneous high-cadence spectra of six EUV lines formed 
at chromospheric, transition region and coronal temperatures, used to determine
spatially resolved intensities and line-of-sight flow speeds at flaring pixels. 
Emission lines of the following ions were selected, which we list together with their
nominal wavelength and formation temperature \citep{1990A&AS...82..229L,2002A&A...381..653S}: He~\textsc{i} ($\lambda = 584.3$~{\AA},
$T\sim3.9\times 10^4$~K), O~\textsc{iii} ($\lambda =599.6$~{\AA}, $T\sim 1.2\times
10^5$~K), O~\textsc{v} ($\lambda= 629.7$~{\AA}, $T\sim 2.6\times 10^5$~K), Ne~\textsc{vi} ($\lambda=562.8$~{\AA},
$T\sim 4.2\times 10^5$~K), Mg~\textsc{ix} ($\lambda= 368.0$~{\AA}, $T\sim 1.0\times10^6$~K)
and Si~\textsc{xii} ($\lambda= 520.7$~{\AA}, $T\sim 2\times 10^6$~K). The CDS slit
covered an area of $2''\times 140''$ with a pixel size of $2''\times 1.7''$. 
The spatial resolution of CDS is larger than this, about $5''$--$6''$, 
due to the size of the point spread function  \citep{1999ApOpt..38.7035P}. 
During our observing campaign, the CDS slit was set fixed in a
sit-and-stare mode (i.e.\ no compensation for solar rotation was applied) 
acquiring observations at a cadence of $\sim$15~s (exposure time 10~s) during the period $\sim$7:25~UT to 12:55~UT. 
Before and after the 1D observing sequence,
four raster scans of 20 successive steps of the CDS slit in
$x$-direction, corresponding to an area of 40$''\times 140''$, were
obtained for co-alignment purposes. 
The CDS data were corrected for the CCD bias,
deviations in exposure time, flat field and cosmic ray hits using CDS
standard calibration software available in the SolarSoftWare (SSW) tree.

CDS spectra acquired after the recovery of the SOHO spacecraft in 1998 are characterized by broad and asymmetric line profiles,
and were fitted with a single broadened Gaussian component\footnote{CDS software note Nr.~53:
http://orpheus.nascom.nasa.gov/cds/swnote/cds\_swnote\_53.ps}.
During the flare impulsive phase, the O~\textsc{v}  profiles in several pixels along the slit were 
composed of more than one component, and were hence fitted with a two component broadened
Gaussian model. The line fits provided the specific peak intensities, the wavelength positions of the core and the width of the line. 
The selected spectral lines all showed preflare and postflare components. 
Of particular interest for our study are the Doppler velocities of mass flows caused by the flare energy deposited 
in the chromosphere, which were calculated from the wavelength shift of the line centroid relative to its nominal wavelength.
Positive velocities (red-shifts) denote motions away from the observer, i.e.\ toward the solar surface; negative velocities (blue-shifts) 
indicate motions upward into the corona. 
The quiet Sun shows characteristic flow patterns, in the transition region dominated by continuous red-shifts.
Since we are interested in the chromospheric evaporation flows induced by the flare energy release, we ``compensate'' for the
continuous quiet Sun flows by determining the reference wavelength (``zero-velocity'') by averaging for each pixel the derived 
center wavelength position during the postflare phase 11:00--12:55~UT. We note that this procedure gives relative velocities 
(i.e.\ not absolutely calibrated). 
In this paper, we concentrate on the evolution in the He~\textsc{i}, O~\textsc{v} and Si~\textsc{xii} lines,
which provide the best signal and cover the full temperature range from the chromosphere to corona. 
The 1$\sigma$-uncertainties of the velocities are estimated to 5~km~s$^{-1}$ for He~\textsc{i},
10~km~s$^{-1}$ for O~\textsc{v} and 20~km~s$^{-1}$ for Si~\textsc{xii}.

On 2006 July 6, we acquired a sequence of MDI full-disk line-of-sight magnetograms \cite[1.96$''$/pixel;][]{scherrer1995} 
with a cadence of 1~min. Each hour, one MDI white light image was also taken.
For the global context of the event as well as for co-alignment purposes, we also used the EUV full-disk images acquired in  
the 304~{\AA} (He~{\sc ii}), 171~{\AA} (Fe~{\sc ix/x}), and 195~{\AA} (Fe~{\sc xii/xxiv}) filters by 
the Extreme-ultraviolet Imaging Telescope \citep[SOHO/EIT;][]{delaboudiniere1995}. 
EIT observed with a time cadence of $\sim$12~min in the 195~{\AA} passband and 6~hours in the other passbands, with a pixel resolution of $\sim$2.6$''$. The CDS rasters were taken close in time to the EIT cycle through all four wavelengths, which is regularly done each 6~hours, to optimize the data co-alignment. 
Dark current and flat field were corrected using the EIT data reduction 
routines available in SSW.

\subsection{TRACE}

The TRACE satellite \citep{handy1999} provided image sequences in the 171~{\AA} (Fe~{\sc ix/x}) passband, observing plasma at a temperature of $\sim$10$^6$~K, with a cadence of $\sim$80~s. Each 10 min, we obtained a cycle of TRACE images at 1216~{\AA} (Hydrogen Ly$\alpha$), 1600~{\AA} (UV continuum), 1550~{\AA} (C~{\sc iv}) and white light. The TRACE filtergrams feature the solar photosphere, chromosphere, transition region and corona with a spatial sampling of 0.5$''$/pixel 
for a $\sim$511$''\times 511''$ field-of-view (FoV). We applied TRACE flat fielding and
dark current subtraction using the SSW TRACE data reduction routines. The effect of cosmic ray particle hits present in some TRACE images
was reduced by median filtering.

\subsection{RHESSI}

The Ramaty High Energy Solar Spectroscopic Imager \citep[RHESSI;][]{lin2002} observes
high energy solar flare emission from 3~keV to 17~MeV with high spectral and spatial resolution
for a full-Sun FoV. RHESSI uses a set of nine rotating modulation collimators consisting of pairs of widely separated X-ray opaque grids 
with high-sensitive Germanium detectors behind each collimator. The X-ray images of the source are reconstructed 
by ground-based software \citep{schwartz2002} from the incident photon fluxes, which are time-modulated by the
nine modulation collimators as the spacecraft rotates at 15 revolutions per minute \citep{hurford2002}.

The impulsive phase of the M2.5 flare of 2006 July 6 was fully captured by RHESSI observations. 
We reconstructed RHESSI images in the 6--12 and 20--60~keV energy 
bands with the CLEAN algorithm using grids 3 to 8, giving a spatial resolution of $\sim$7$''$
\citep{hurford2002}. In the 6--12~keV band (dominated by thermal emission of plasma with temperatures $\gtrsim$$10^7$~K), we reconstructed images for 
consecutive intervals of 30~s. In the 20--60~keV band (dominated by nonthermal emission of electron beams), 
we reconstructed images over individual bursts in the RHESSI high-energy
light curves with integration intervals in the range 30 to 60~s. 
In addition, for the determination of footpoint source sizes, we also reconstructed RHESSI images over selected 
peaks in the 20--60~keV energy band with the Pixon and the MEM-NJIT algorithms using grids 1 to 8. 
Pixon and MEM-NJIT are suitable algorithms to provide reliable estimates of the reconstructed X-ray source sizes \citep{2009ApJ...698.2131D}. 
Note that for these image reconstructions during the HXR peaks, we also used the finest RHESSI grid no.~1 (only possible in case of good count statistics), 
which has a FWHM angular resolution of 2.3$''$. 
RHESSI spectra were extracted with 1~keV spectral resolution using all front detectors 1 to 9 except 2 and 7 (with lower
spectral resolution and high threshold energies), deconvolved with the full detector response matrix \citep{smith2002} and fitted with an isothermal plus thick-target bremsstrahlung spectrum \citep{holman2003} for consecutive 12~s 
intervals throughout the flare impulsive phase.

\subsection{Groundbased H$\alpha$ imaging}

During JOP171, we acquired full-disk H$\alpha$ images at Kanzelh\"ohe Observatory with a cadence of $\sim$3~s. The images have a spatial sampling of 2.2$''$/pixel and are tracked and aligned using a solar limb-finding algorithm \citep{otruba03}. In addition, we acquired H$\alpha$ filtergrams of AR 10898 at 
Hvar Observatory \citep{otruba05} with a cadence of $\sim$4~s. The Hvar H$\alpha$ images cover a FoV of about $300'' \times 300''$ centered on the AR, 
and were co-aligned via cross-correlation techniques with Kanzelh\"ohe H$\alpha$ images.

\subsection{Data co-alignment}

Much care has to be taken in such an extensive multi-wavelength study combining imaging and 
spectroscopic observations from various instruments, to properly co-align 
all the different data sets. Such considerations were  
already part of the planning of the JOP171 observing run. 
The co-alignment between different instruments was derived
by the two-dimensional cross-correlation of images or CDS raster scans
sensitive to similar temperature ranges (showing similar atmospheric layers) and recorded close in
time to each other. Generally, several pairs of images/scans were used for the cross-correlations, and a
linear approximation of the shifts in time led to offsets
of the coordinate systems with uncertainties typically below $\sim$1$''$.
The reference coordinate system for our data set was EIT, which has 
a pointing accuracy of $\sim$2$''$. First, CDS He~\textsc{i} 584~{\AA} raster images were
co-aligned with the EIT He~\textsc{ii} 304~{\AA} full disk image, which was taken roughly at the same time.
TRACE 171~{\AA} maps were co-aligned with the EIT maps in the same
wavelength band, and the pointing information was used to update all
other TRACE filtergrams, accounting for the varying FoV of the individual TRACE filters. The co-aligned
TRACE white light images were the reference for the MDI full-disk white light images, and the determined offsets were also applied to the
MDI magnetograms. 
Kanzelh\"ohe full-disk H$\alpha$ images were co-aligned with 
EIT 195\,\AA~images of the active region. 
Hvar H$\alpha$ maps were then co-aligned 
with the Kanzelh\"ohe H$\alpha$ images. RHESSI has a pointing accuracy of better than 1$''$, and
the pointing information given in the image fits header was not altered.
With the resulting co-alignment, brightenings along the CDS slit coincided well with bright flare kernels 
observed in H$\alpha$ and in TRACE UV images, which further coincided with the 
RHESSI HXR footpoint emission observed at several time steps. 
We also quantified the errors in our co-alignment by repeating the procedure at different times and in changing 
the sequence of the instruments and filters. We estimate that 
the co-alignment for our data set is correct to within $\sim$2$''$.

\section{Results}

\subsection{Event overview and context observations}

\begin{figure}
\resizebox{7cm}{!}{\includegraphics[clip=true]{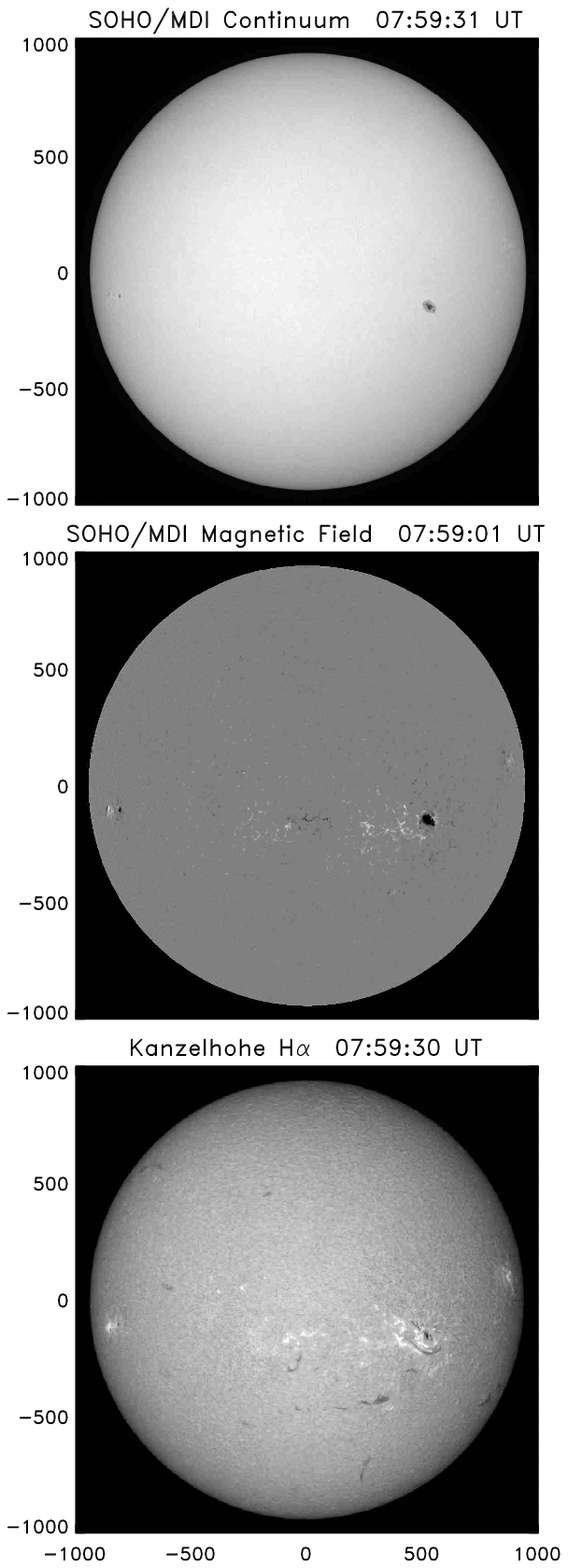}} 
\caption{Overview of AR 10898 (S06, W34) on 2006 July 6, shortly before the start  of the M2.5 flare. From top to bottom: SOHO/MDI continuum image, SOHO/MDI
longitudinal magnetic field map, Kanzelh\"ohe Observatory H$\alpha$ filtergram. The units on the $x$- and $y$-axis are in arcseconds; the same holds for all images 
plotted in the following.
 } \label{fig_overview}
\end{figure}

\begin{figure}
\resizebox{12cm}{!}{\includegraphics[clip=true]{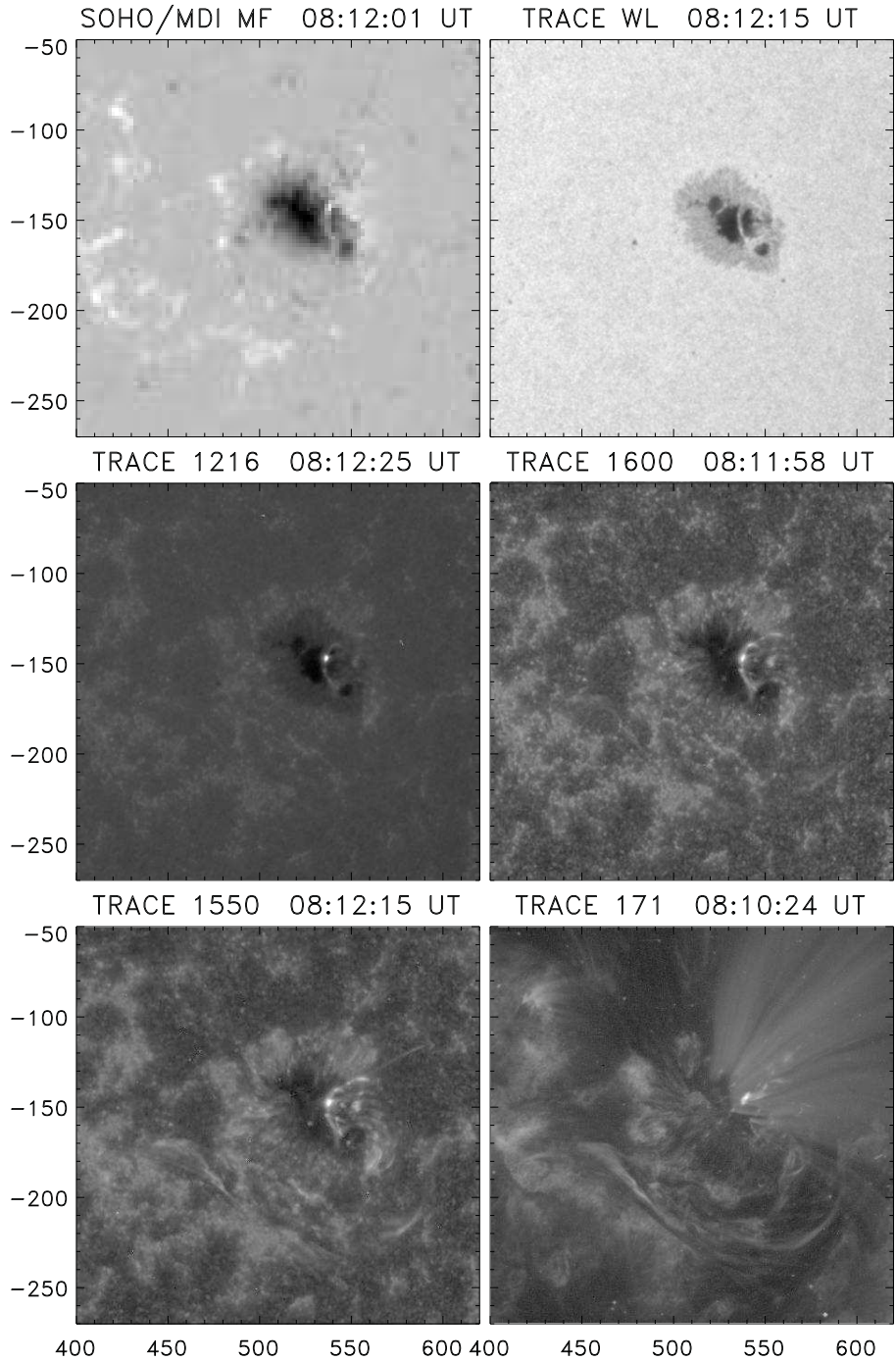}} 
\caption{Multiwavelength imaging at the beginning of the flare: SOHO/MDI longitudinal magnetic field, TRACE white light, TRACE 1216, 1550, 1600, and 171~{\AA} filtergrams. 
In the accompanying movie1, we show the evolution of the event in the TRACE 171~{\AA} passband.
 } \label{fig_overview_i6}
\end{figure}

On 2006 July 6 the instruments involved in the JOP171 campaign pointed to
AR~10898 located at 6$^\circ$ south and 34$^\circ$ west from the solar meridian. Figure~\ref{fig_overview} shows a full-disk white light image and 
longitudinal magnetogram from SOHO/MDI together with an H$\alpha$ filtergram from Kanzelh\"ohe Observatory taken directly before the onset of the M2.5 flare. Figure~\ref{fig_overview_i6} shows a subfield around AR 10898 observed by MDI 
and the different TRACE passbands. The active region appears rather simple (Wilson classification $\beta$), 
with the negative polarity leading sunspot embedded in a network cell of predominantly positive polarity fields. The white light image reveals two light bridges across the sunspot, the most prominent one crossing the spot in north-south direction. 

Figure~1  in \cite{2009A&A...505..811B} shows a high-resolution view of AR 10898 obtained on 2006 July 4 with the Dutch Open Telescope in G-band, Ca~{\sc ii}~H and H$\alpha$ together with a high-resolution MDI magnetogram, a TRACE 171~{\AA} and TRACE white light frame. 
The MDI high-resolution magnetogram acquired on July 4 shows that the 
magnetic elements surrounding the sunspot are highly intermixed on small scales. It can also be seen that early on that day, the light bridge was not yet formed. It started forming later on the day of July 4, associated with a distinct anti-clockwise rotation of the sunspot between 2006 July 4 and 6. 
We note that AR 10898 did not produce flares above B-level from June 26, when it rotated over the Eastern solar limb, until July 4. On July 4, it was the source of several microflares \cite[studied in detail in][]{2009A&A...505..811B} and a C1-class flare. On July 5, it produced 2 C-class flares,
increasing to its maximum flare activity on July 6, where it was the source of the M2.5 flare under study. 

The M2.5 flare that occurred on 2006 July 6 around 8~UT was a textbook two-ribbon flare with associated filament eruption and coronal mass ejection. 
The associated CME was fast ($\sim$900 km~s$^{-1}$ in the SOHO/LASCO coronagraph's FoV). Its main acceleration phase was studied in detail in \cite{temmer2008}, revealing a peak acceleration of 1.1~km~s$^{-2}$ around 08:20~UT. The event was accompanied by a propagating shock wave, as is evidenced by the coronal and interplanetary type~II burst 
observations of the AIP (Astrophysikalisches Institut Potsdam) and  Wind/WAVES dynamic radio spectra. Finally, we note that
the event was associated with distinct, long-lasting bipolar coronal dimming regions (see movie1 for the evolution in TRACE 171~{\AA}), which are generally interpreted as being due to mass depletion in the wake of the erupting CME and associated field line opening \citep[e.g.][]{1996ApJ...470..629H,1999ApJ...520L.139Z}. The evolution of this dimming region was studied by several groups \citep{2007ApJ...660.1653M,2007ApJ...662L.131J,2008SoPh..252..349A}.

\subsection{Multi-wavelength imagery of the M2.5 flare}

\begin{figure}
\resizebox{15cm}{!}{\includegraphics[clip=true]{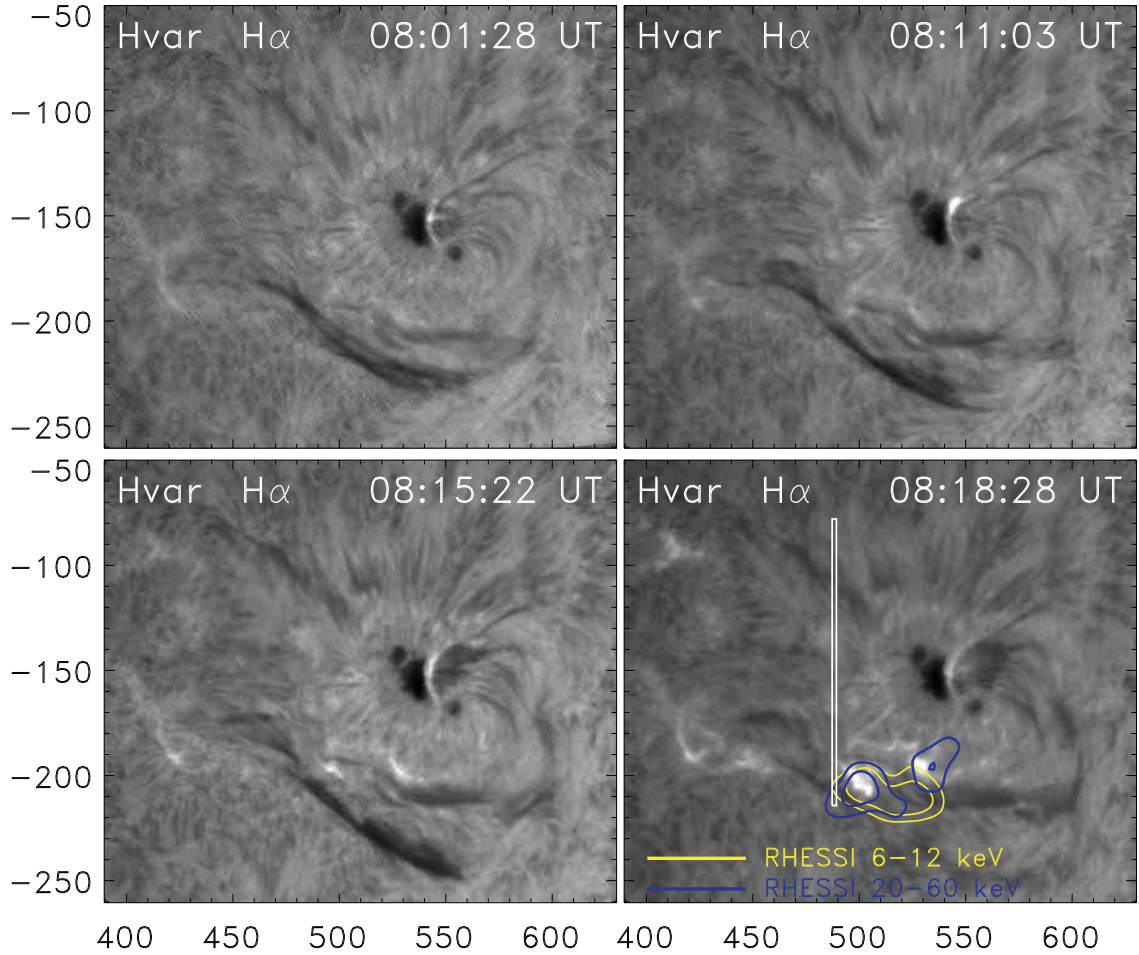}} 
\caption{Early phase of the flare as observed in the H$\alpha$ spectral line at Hvar Observatory. Note the brightenings in the light bridge (top panels)
and associated mass flows from the light bridge (bottom panels). Blue and yellow contours in the last panel are RHESSI 6--12 keV (thermal) and 20--60 keV 
(nonthermal) flare radiation. The vertical rectangle indicates the position of the CDS slit.
 } \label{fig_hvar}
\end{figure}

\begin{figure}
\resizebox{12cm}{!}{\includegraphics[clip=true]{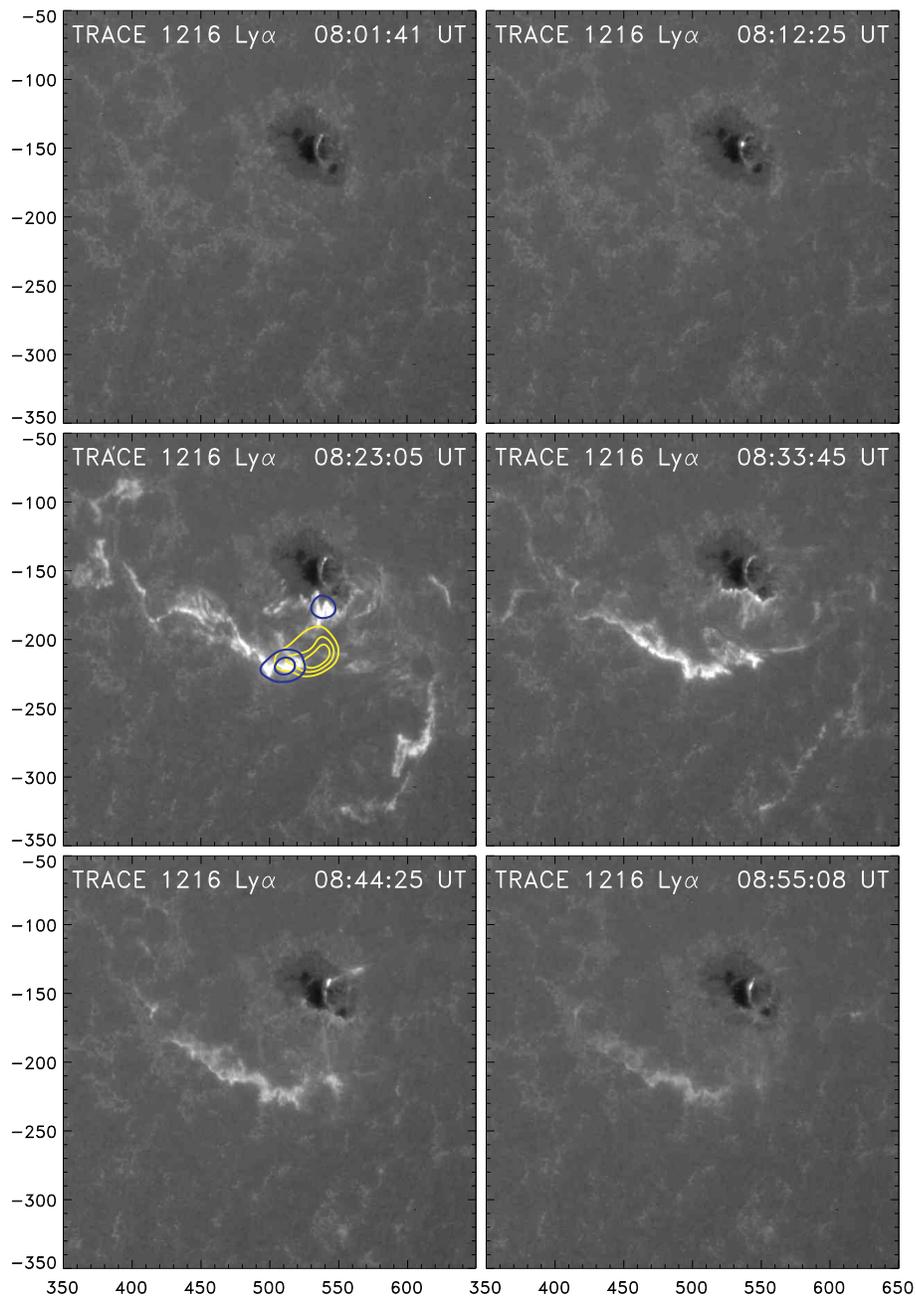}} 
\caption{Evolution of the flare and the brightenings in the sunspot's light bridge as observed in TRACE 1216~{\AA} 
Hydrogen Ly$\alpha$ spectral line from 08:01 to 08:55~UT. Images are logarithmically scaled. The contours in the third panel are co-temporal RHESSI 20--60~keV 
(blue; 30\% and 70\% of the image's maximum)  and 6--12 keV emission (yellow; 40, 60 and 80\% of the image's maximum). 
In the accompanying movie2, the full time range from 07:50 to 12:50~UT 
is presented, showing the ongoing brightenings and activations in the light bridge, before, during, and after the flare. 
 } \label{fig_trace_1216}
\end{figure}

The M2.5 flare of 2006 July 6 is preceded by activation of the AR filament (starting as early as 07:40~UT) and by distinct localized brightenings and mass flows in the sunspot light bridge observed in TRACE 1216, 1550, 1600, 171~{\AA} and H$\alpha$ images around 08:01 and 08:11~UT (see Figs.~\ref{fig_overview_i6} and \ref{fig_hvar}). We note that continuous brightenings and associated mass flows in the light bridge can be observed during the overall JOP171 observing period on July 6 from 07:25 to 12:55~UT. The brightenings appear most pronounced in the TRACE Ly$\alpha$ 1216~{\AA} spectral line (see Fig.~\ref{fig_trace_1216} and movie2). 
Light bridges are regions of strong magnetic field gradients within a sunspot, and may thus be favorable locations for current sheets and magnetic reconnection to occur 
\cite[e.g.][]{1995A&A...302..543R,1997ApJ...484..900L,2002ApJ...565.1335Q}. 
Recently, \cite{2009ApJ...696L..66S} studied Hinode/Solar Optical Telescope high-resolution vector magnetic field measurements 
of a light bridge that produced recurrent chromospheric plasma ejections. These authors report the possible direct detection of electric 
currents flowing in the current sheet formed at the magnetic reconnection sites above the light bridge.

\begin{figure}
\resizebox{12cm}{!}{\includegraphics[clip=true]{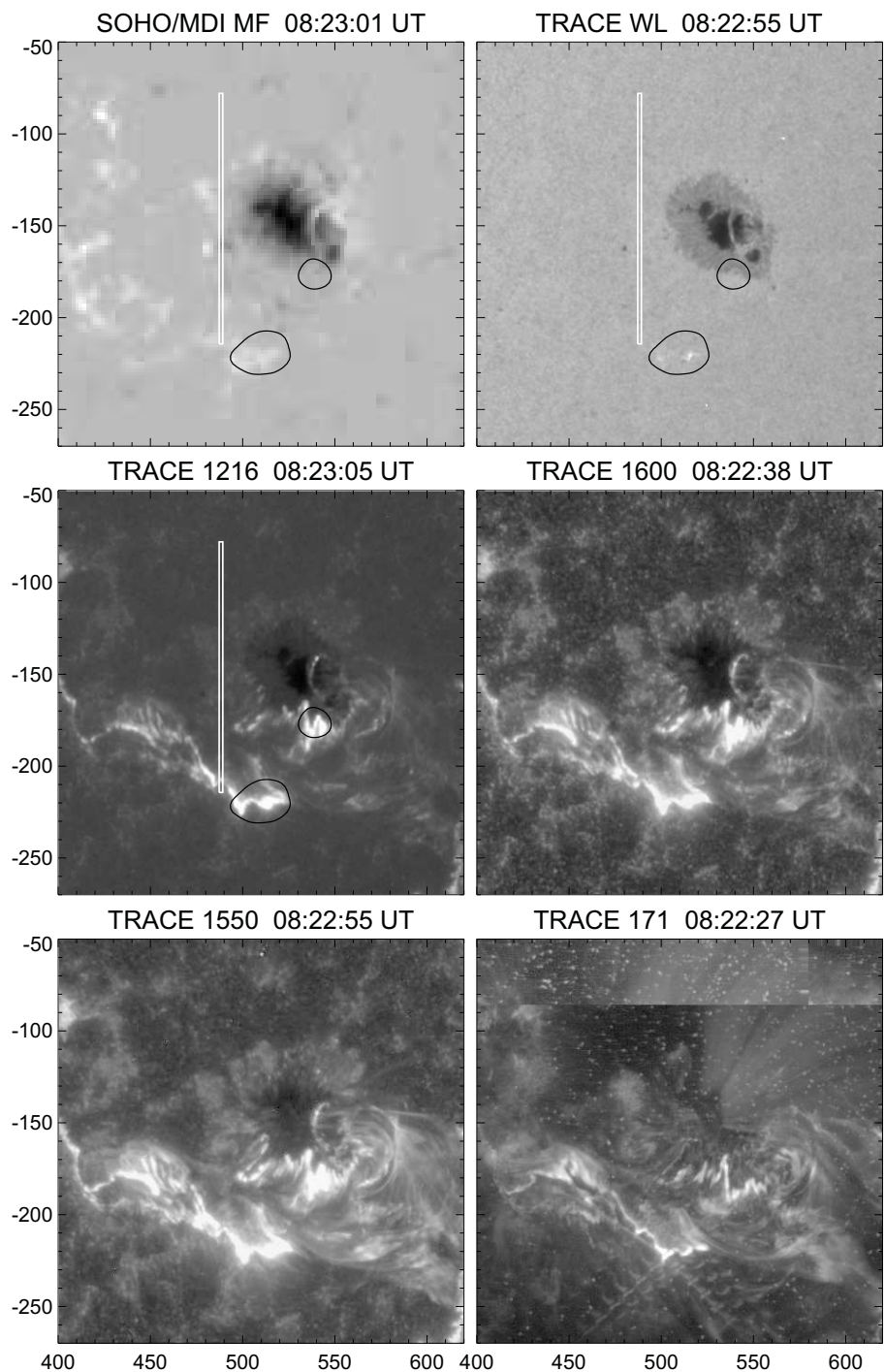}} 
\caption{Multiwavelength imaging during the flare maximum: SOHO/MDI longitudinal magnetic field, TRACE white light, TRACE 1216, 1550, 1600, and 171~{\AA} filtergrams. 
The white rectangle outlines the CDS slit. The contours indicate the contemporaneous RHESSI 20--60 keV hard X-ray emission at 30\% of the image's maximum. 
Note the TRACE white light enhancement which coincides with the southern hard X-ray footpoint observed by RHESSI. 
 } \label{fig_overview_max_i7}
\end{figure}

The impulsive phase of the M2.5 flare under study starts with two footpoint brightenings south of the main sunspot of AR~10898, observed in H$\alpha$ and in RHESSI HXRs around 08:15--08:18~UT (Fig.~\ref{fig_hvar}), which then develop to extended flare ribbons expanding in a north-south direction. In Figure~\ref{fig_overview_max_i7}, we show images obtained in the different TRACE passbands around the flare maximum. It can be seen that in its maximum phase, the flare is associated with distinct enhancements 
in white light at the southern flare ribbon. These white light enhancements spatially coincide with the RHESSI 20--60~keV footpoint emission from flare-accelerated electrons. 

In Figure~\ref{fig_trace_rhessi}, we show a sequence of the flare during the impulsive phase up to its decay phase observed in the TRACE 171~{\AA} passband together with RHESSI 6--12 and 20--60~keV X-ray emission (see also online movie3). The 20--60~keV nonthermal HXR emission is located along the strongest brightenings of the TRACE flare ribbons. 
The 6--12~keV emission comes from the region between the RHESSI HXR footpoints, located above the TRACE postflare loops, evidencing hot plasma ($T \sim 20$~MK)
from the rising flare loop system. In the accompanying movie no.~3 one can see that the TRACE 171~{\AA} ($T \sim 1$~MK) postflare loops first appear in the image at 
08:35:50~UT, whereas the hot RHESSI flare loops can be observed already around 08:18~UT. This time difference of 18~min provides a rough 
estimate of how long it takes the hot flaring plasma to cool from about 20~MK to 1~MK, which is predominantly due to conductive cooling \citep[cf.\ discussions in][]
{2006SoPh..234..273V}.

\subsection{CDS spectroscopy}

\begin{figure}
\resizebox{10cm}{!}{\includegraphics[clip=true]{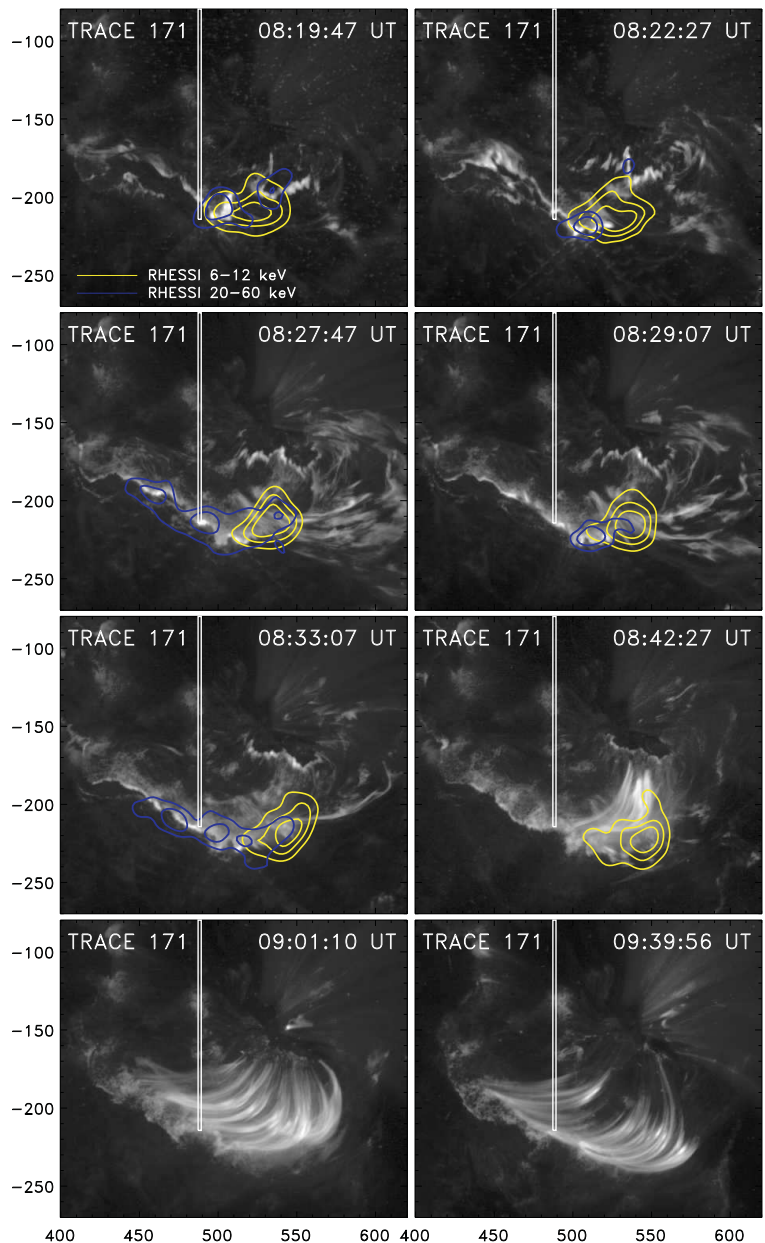}} 
\caption{Sequence of TRACE 171~{\AA} images together with contours of RHESSI 6--12 (yellow) and 20--60 keV (blue) images closest in time. 
Contour levels are 40, 60 and 80\% of each image's maximum in the 6--12 keV energy range; 50\% and 75\% levels in the 20--60~keV range. At 08:45 UT 
the RHESSI spacecraft entered the Earth shadow stopping the observations. 
The accompanying movie3 shows the flare evolution in TRACE 171~{\AA} together with RHESSI 6--12 and 20--60 keV contours during the impulsive phase.
 } \label{fig_trace_rhessi}
\end{figure}

\begin{figure}
\resizebox{14cm}{!}{\includegraphics[clip=true]{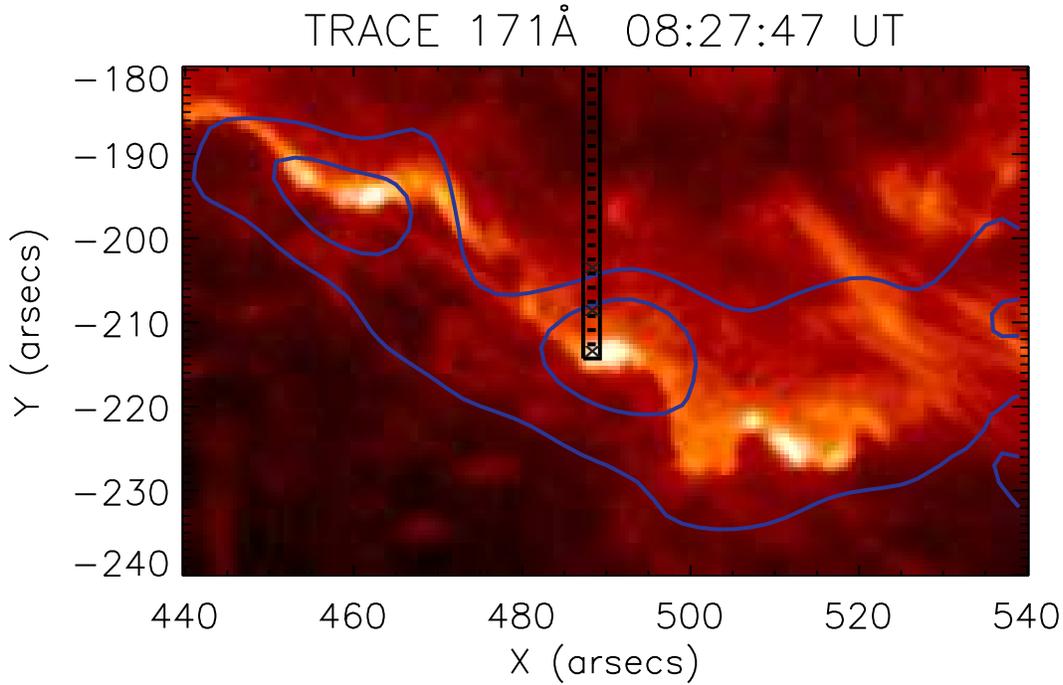}} 
\caption{Small subfield of a TRACE 171~{\AA} image (taken at 08:27:47~UT) together with contours of the RHESSI 20--60 keV image (blue) closest in time. 
The black rectangle shows the location of the CDS slit, indicating the size of the individual CDS pixels by the horizontal lines. 
Roughly ten CDS pixels located at the bottom part of the slit are at some instant located along the flare ribbon (compare also Fig.~\ref{fig_trace_rhessi}).
The crosses mark those CDS pixels whose evolution is shown in Figs.~\ref{fig_cds_long}, \ref{fig_cds_short} and \ref{fig_double_comp}.
 } \label{fig_trace_cds}
\end{figure}

\begin{figure}
\resizebox{16cm}{!}{\includegraphics[clip=true]{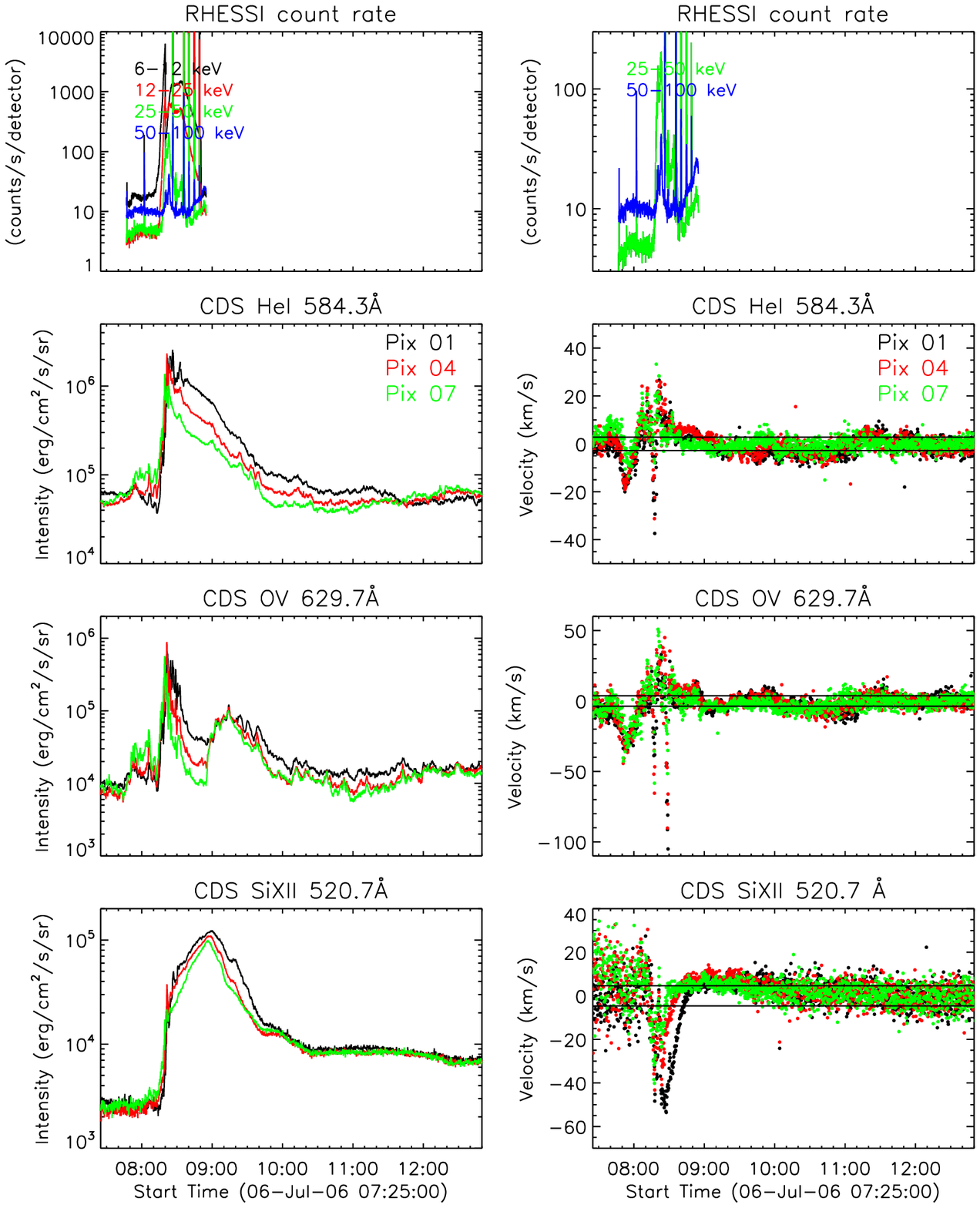}} 
\caption{Evolution of integrated intensities (left; plotted on a logarithmic scale) and velocities (right) derived from the CDS spectral fits in He\,{\sc i}, O\,{\sc v} 
and Si\,{\sc xii} for the full CDS observation period in three different pixels along the lower end of the CDS slit (marked by crosses in Fig.~\ref{fig_trace_cds}). 
Horizontal lines in the velocity curves indicate the 1$\sigma$-uncertainties in the velocity calibration (derived during the postflare phase 
11:00 to 12:55~UT).  } \label{fig_cds_long}
\end{figure}

Fig.~\ref{fig_trace_rhessi} shows a sequence of TRACE 171~{\AA} images together with the position of the CDS slit, revealing that 
the slit crosses the southern flare ribbon. Roughly 10 pixels along the lower end of the CDS slit are at some instant located directly at the flare ribbon, 
which moves southward as the flare progresses. In Fig.~\ref{fig_trace_cds}, we plot a small TRACE 171~{\AA} subfield around the southern flare ribbon 
together with the RHESSI 20--60~keV contours and the CDS slit, indicating the size of the individual CDS pixels. 
In Fig.~\ref{fig_cds_long} we show the evolution in three selected pixels for the full CDS observing period. We plot the 
integrated intensity and the velocity derived from the one-component Gaussian fits to the CDS spectra in the He\,{\sc i}, O\,{\sc v} and Si\,{\sc xii} lines.
The 1$\sigma$-uncertainties of the velocity calibration derived from the CDS spectra in the postflare phase 
are indicated by horizontal lines.

\begin{figure}
\resizebox{12.cm}{!}{\includegraphics[clip=true]{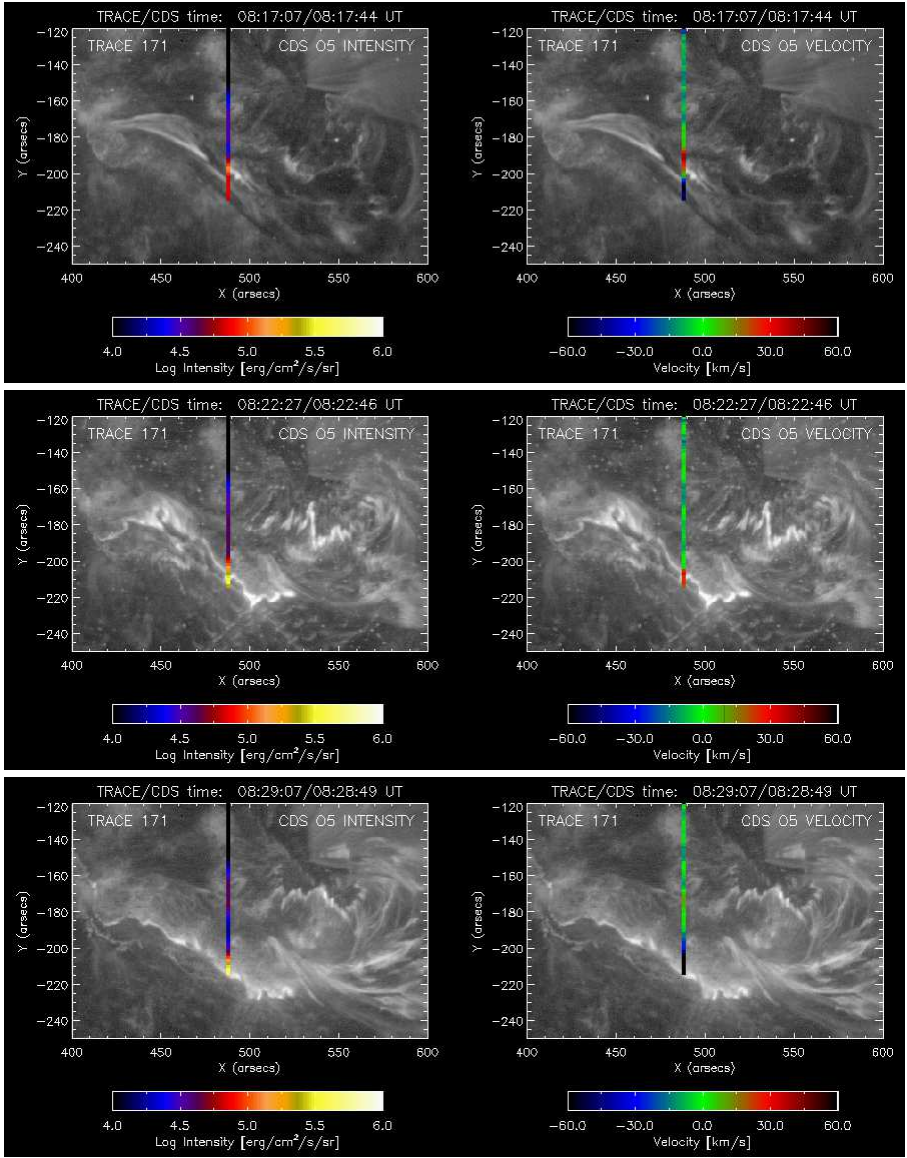}} 
\caption{Three snapshots of TRACE 171~{\AA} images combined with the integrated intensities (left) and velocities (right) derived from the Gaussian fit to a co-temporal 
CDS spectrum in the O\,{\sc v} line. The top panel is taken during the time of the filament lift-off and commencement of the impulsive flare brightenings 
(the CDS velocities are blue-shifted over the filament, and red-shifted at the flare kernels), the middle panel around the time of the highest HXR peak (showing distinct red-shifts at the CDS pixels crossing the flare ribbons), and the bootom panel during the time of the distinct late RHESSI peak 
associated with the strongest transition region upflows at the CDS pixels crossing the flare ribbons.
The whole evolution is shown in the accompanying movie4 (note that only each second CDS spectrum available is plotted in the movie).
 } \label{fig_snapshot}
\end{figure}

Fig.~\ref{fig_cds_long} reveals three distinct periods with upflows observed in the O\,{\sc v} line. The first two peaks, 
around 07:45--08:05~UT and 08:15--08:20~UT are related to the activation and eruption of the AR filament. 
This can be best confirmed and followed in the online movie4, where we combine the imaging information from TRACE 171~{\AA} images 
with integrated intensities and velocities observed in O\,{\sc v} along the CDS slit. 
Figure~\ref{fig_snapshot} shows three snapshot of this movie. The top panel is taken during the time of filament lift-off and the commencement of the impulsive flare phase.
It reveals blue-shifts in the pixels covering the rising filaments, and red-shifts (due to chromospheric evaporation downflows) in the pixels covering the bright flare footpoints. We also note that at some instants during the filament activation and lift-off several CDS pixels covering the eruption appear red-shifted
(see movie4). This may be an effect of internal motion of the filament (twisting) while it rises.
The filament lift-off is observed in all three spectral lines during 08:15--08:20~UT with line-of-sight 
velocities in the CDS FoV up to 40--60~km~s$^{-1}$, indicating 
the multithermal nature of the erupting filament plasma covering a broad temperature range from $\sim$$10^4$ up to at least 
$2 \times 10^6$~K.

\begin{figure}
\resizebox{12cm}{!}{\includegraphics[clip=true]{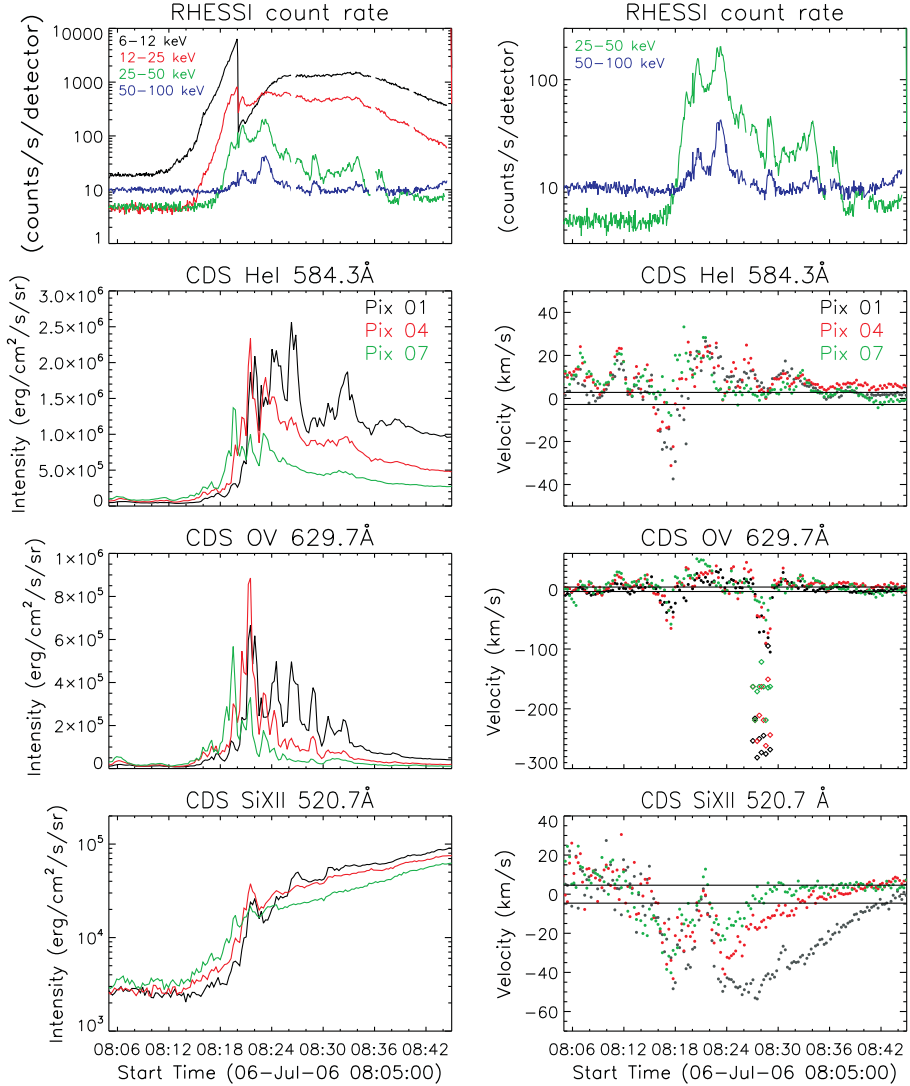}} 
\caption{Flare impulsive phase. First row: RHESSI X-ray flux at several energy bands from 6 to 100 keV. (The discontinuity at the low energy bands 
at 08:20~UT is an artifact due to the change in the RHESSI attenuator state from A0 to A1.) Second to
fourth row: Evolution of integrated intensities (left) and velocities (right) derived from the one-component Gaussian fits 
to the He\,{\sc i}, O\,{\sc v}, and Si\,{\sc xii} CDS spectra
during the flare impulsive phase in three different pixels along the lower end of the CDS slit (marked by crosses in Fig.~\ref{fig_trace_cds}). Horizontal lines in the velocity curves indicate the 1$\sigma$-uncertainties in the velocity calibration.
For the O\,{\sc v} line, we also plot the velocities derived from the two-component 
fits during the period 08:27--08:29~UT (indicated by triangles). Note that the He\,{\sc i} and O\,{\sc v} intensities are plotted
on a linear scale, to make the bursty nature clearer, whereas in Fig.~\ref{fig_cds_long} they are plotted logarithmically.
 } \label{fig_cds_short}
\end{figure}

Figure~\ref{fig_cds_short} shows the CDS spectroscopy results together with the RHESSI X-ray flux at different energy bands for the 
flare impulsive phase. The integrated intensities in He\,{\sc i} and O\,{\sc v} show impulsive behavior with a 
sequence of distinct peaks (with a typical duration of $\sim$1~min), 
whereas the Si\,{\sc xii} intensities show only one peak around 08:21~UT, coincident with peaks in He\,{\sc i} and O\,{\sc v}, but a gradual increase thereafter.
It is worth noting that the intensity peaks observed in adjacent CDS pixels do not necessarily occur at the same 
time and with the same strength, due to the flare spatial/temporal evolution along the CDS slit. 
A similar argument holds when we compare the CDS intensities in the chromospheric 
(He\,{\sc i}) and transition region lines (O\,{\sc v}) with the RHESSI HXR flux: the peaks are somehow related but do not occur simultaneously. 
This can be explained by the fact that in the RHESSI light curves we observe the spatially integrated X-ray flux, whereas in the CDS light curves constructed for individual pixels, we see the emission from a certain (small $2'' \times 1.7''$) subfield of the flaring region at a certain instant.

The strongest HXR peaks occur between 08:20--08:26~UT, with the increase of the nonthermal emission ($\gtrsim$20~keV) 
starting already around 08:18~UT 
(see also the top panel in Fig.~\ref{fig_cds_short}). We note that during this 
period, the CDS slit is not directly located on the site of strongest energy deposition as evidenced by the RHESSI HXR footpoints 
but crosses the flare ribbon slightly to the East of the HXR footpoints (cf.\ Figs.~\ref{fig_overview_max_i7} and \ref{fig_trace_rhessi}). 
During this period, we also observe a sequence of strong peaks in the integrated intensities of the He\,{\sc i} and O\,{\sc v} lines, indicating that
the CDS slit is located on energized parts of the flare ribbons, as well as plasma downflows in the chromospheric and transition region layers
(Fig.~\ref{fig_cds_short}). These downflows reach velocities up to about 20~km~s$^{-1}$ in He\,{\sc i} and 30 to 40~km~s$^{-1}$ in O\,{\sc v}.
In the coronal Si\,{\sc xii} line we observe line-of-sight upflow velocities up to $-50$~km~s$^{-1}$, starting at 08:22~UT and then developing gradually.  
Unfortunately, we do not have a hot flare line formed at temperatures around $10^7$~K to study the behavior of the hottest component of the flaring plasma.
The observed flow behavior (downflows in the chromosphere and lower transition region; upflows in coronal lines) is suggestive of
explosive chromospheric evaporation. This is also supported by the strong HXR peaks at this time, which are due to nonthermal electrons 
impacting on the lower atmospheric layers from a coronal acceleration site, and the associated strong peaks in He\,{\sc i} and O\,{\sc v}.

\begin{figure}
\epsscale{0.95} \plotone{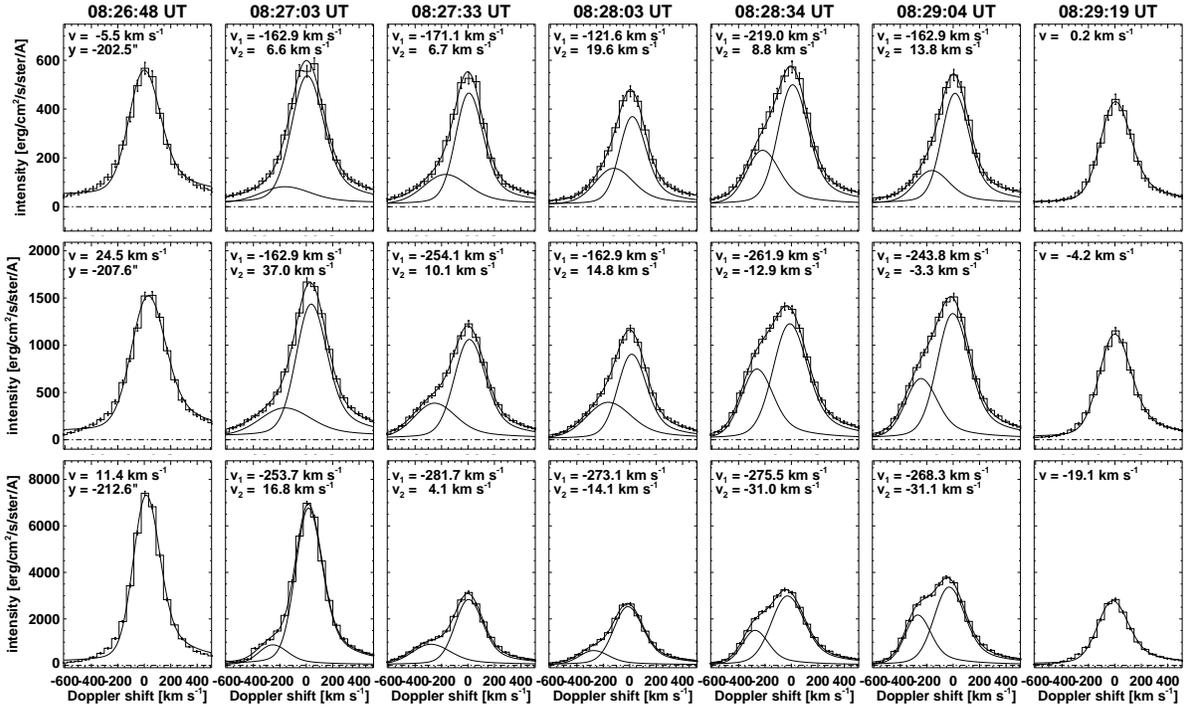}
\caption{Sequence of O\,{\sc v} spectra during the flare peak, fitted with a double component (except for the first and last spectrum of the series) 
for three pixels along the lower end of the CDS slit (marked by crosses in Fig.~\ref{fig_trace_cds}). Only each second spectrum available in the time series is plotted.
 } \label{fig_double_comp}
\end{figure}

We observe another distinct peak in RHESSI HXRs around 08:29~UT. At this time, the CDS slit is crossing the strongest energy deposition site
along the southern flare ribbon, as is evidenced by the RHESSI 20--60 keV images as well as by the bright TRACE flare kernel 
within the spectrometer slit (see Fig.~\ref{fig_trace_cds}). At this time, we also observe distinct peaks in the 
CDS O\,{\sc v} intensity curves in the selected CDS pixels and strong upflows with velocities up to
$-100$~km~s$^{-1}$ derived from the one-component Gaussian fits (Fig.~\ref{fig_cds_short}). However, in 8 pixels on the lower end of the CDS slit, 
we observe double components in the O\,{\sc v} spectra roughly at this period (08:27:03 to 08:29:04~UT). 
In Figure \ref{fig_double_comp} we show the evolution of  O\,{\sc v} spectra in three selected CDS pixels 
(the same pixels, for which we show the intensities and velocities in Fig.~\ref{fig_cds_short}) 
which change from single to double component spectra, together with the single/double component Gaussian fits. 
This spectral evolution is consistent with the bulk of the emission coming from a stationary plasma volume
with an increasing contribution from upflowing plasma elements with velocities in the range of about $-150$ to $-280$~km~s$^{-1}$. (The velocities of the first component lie all within the 3$\sigma$ uncertainties, which we derived for the CDS O\,{\sc v} one-component fits; 
these uncertainties are even larger for the double component fits. Thus we interpret the velocity derived for this component as consistent with emission from a stationary plasma.)
The upflow velocities obtained from the two-component fits are also plotted in Fig.~\ref{fig_cds_short} 
together with the velocities derived from the one-component fits.

\subsection{RHESSI X-ray spectroscopy and imaging}

\begin{figure}
\resizebox{11cm}{!}{\includegraphics[clip=true]{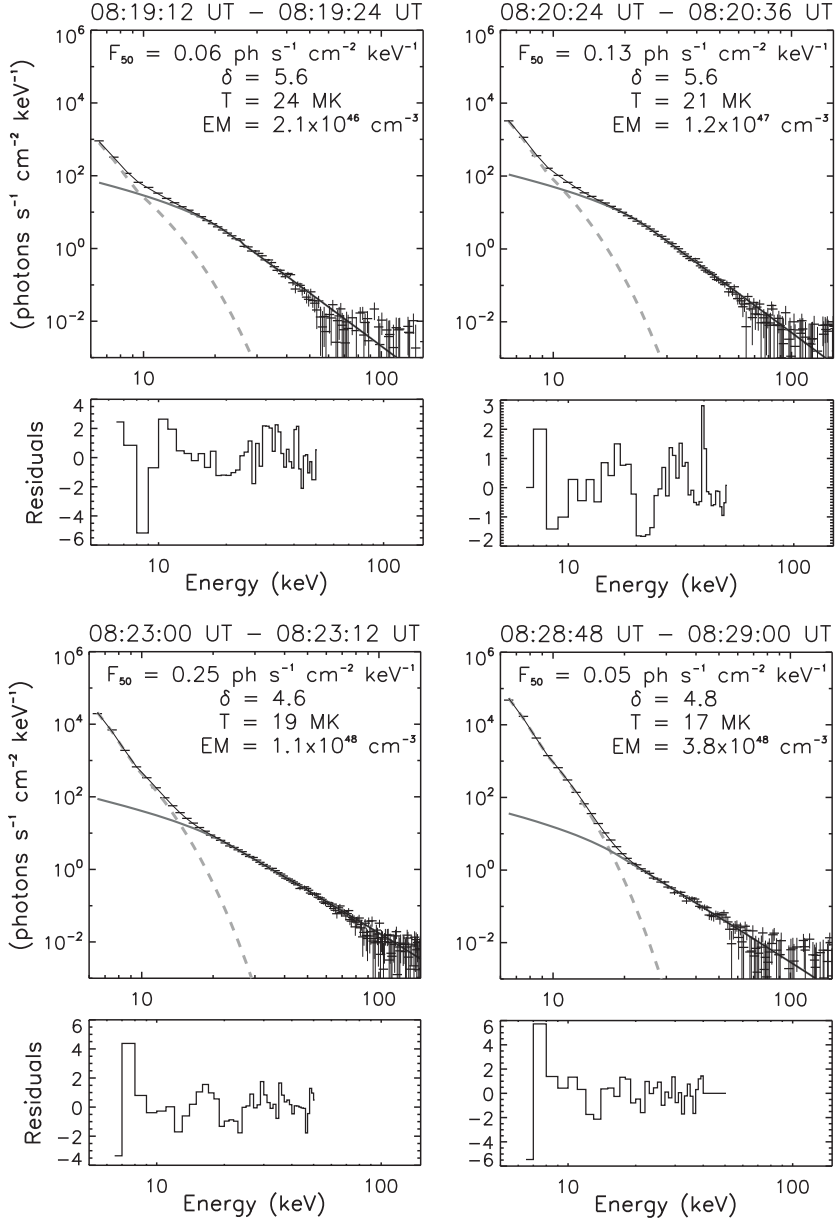}} 
\caption{RHESSI X-ray spectra accumulated during four selected peaks in the RHESSI HXR light curves
(indicated in the time series in Fig.~\ref{fig_rhessi_params}). The top panels show the observed X-ray spectra
together with the two-component fits consisting of the bremsstrahlung and line emission of an 
isothermal plasma (dominating at low energies) plus the nonthermal thick-target bremsstrahlung emission of 
a power-law electron distribution with a low-energy cut-off (dominating at high energies, $\gtrsim$20~keV).
The fit parameters derived are annotated in the spectrum plots.
The bottom panels show the fit residuals (i.e.\ the difference between the observed counts and the model predicted counts 
divided by the estimated 1$\sigma$-uncertainty in the counts) in the fit range. 
 } \label{fig_rhessi_spec}
\end{figure}

\begin{figure}
\resizebox{9.5cm}{!}{\includegraphics[clip=true]{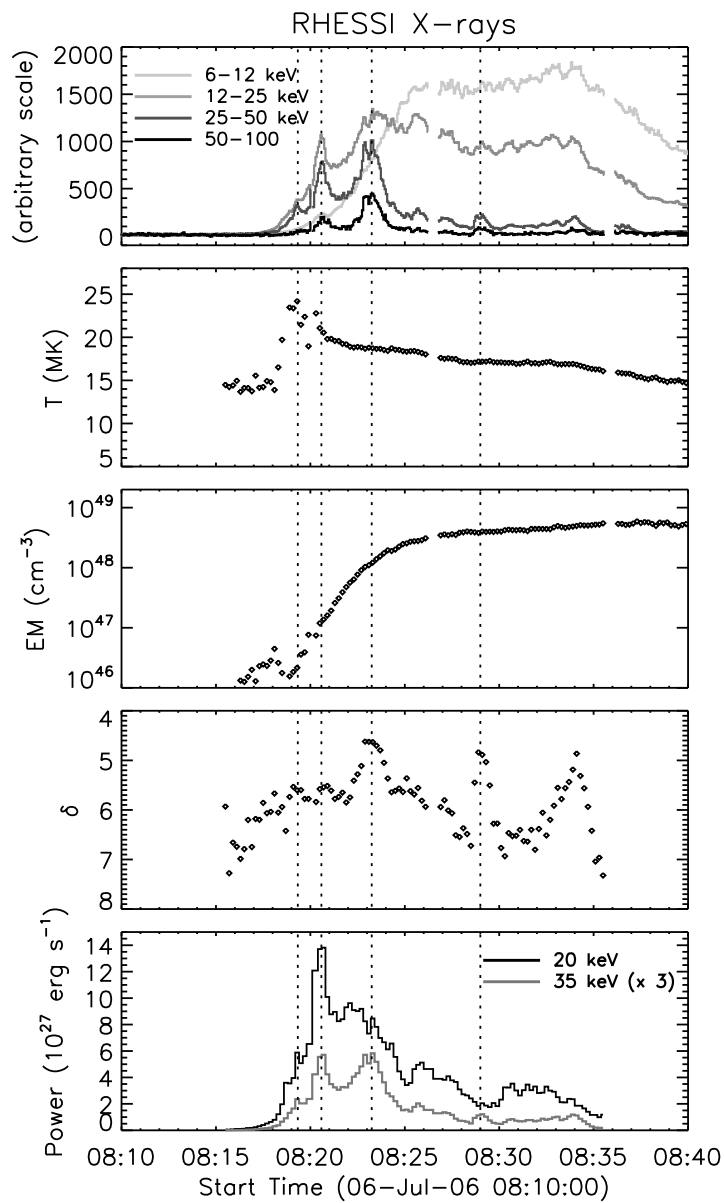}} 
\caption{Time evolution of the thermal plasma parameters and nonthermal electron beam parameters 
derived from RHESSI X-ray spectral fits 
of consecutive 12~s integration intervals. From top to bottom:
Corrected RHESSI count rates in different energy bands (the scales in the different energy bands are altered in order to better see the time evolution), 
plasma temperature, emission measure, electron spectral index~$\delta$, and power in electrons
derived for thick-target emission with cut-off energies of 20~keV and 35~keV, respectively.
The vertical lines denote the time of the spectra plotted in Fig.~\ref{fig_rhessi_spec}.
 } \label{fig_rhessi_params}
\end{figure}

Figure~\ref{fig_rhessi_spec} shows RHESSI X-ray spectra together with the thermal plus nonthermal fits accumulated during four selected 
peaks in the X-ray light curves (indicated by vertical lines in Fig.~\ref{fig_rhessi_params}). 
From these fits, we derive the temperature and emission measure of the hot flaring plasma as well as 
the spectral slope and power in electrons above a cut-off energy~$E_c$ for thick-target emission \citep{1971SoPh...18..489B}. 
The observed RHESSI spectra (e.g., Fig.~\ref{fig_rhessi_spec}) 
indicate that the emission $\gtrsim$20~keV is dominated by nonthermal power-law emission. This is also consistent with
the (upper) estimates of $E_c$ derived from the thick-target fits, which are in the range 15 to 25~keV. Therefore, in the following  we use
a fixed value of $E_c = 20$~keV to determine the kinetic energy of the fast electrons.

Figure~\ref{fig_rhessi_params} shows the time evolution of the plasma and electron beam parameters obtained from the fits to
the RHESSI spectra integrated over consecutive 12~s intervals. From top to bottom we plot RHESSI X-ray lightcurves, plasma temperature, 
plasma emission measure, electron spectral index, and power in electrons for a low-energy cut-off $E_c = 20$~keV and for comparison also for 
$E_c =35$~keV. The temperature peaks already at the first (small) HXR burst around 08:19~UT, with a maximum value of 23~MK. Emission measure is steeply growing
until the largest HXR peak at 08:23~UT and then gradually increasing beyond the end of the enhanced RHESSI HXR emission. 
Such behavior of emission measure and temperature indicates that first a small volume of coronal plasma is 
heated to high temperatures, and subsequently more and more loops are filled with hot plasma. 
The electron spectra are comparatively steep, with the spectrum being hardest at the time of the largest RHESSI peak at 08:23~UT 
with an electron spectral index $\delta \sim 4.5$. 

\begin{figure}
\resizebox{16.5cm}{!}{\includegraphics[clip=true]{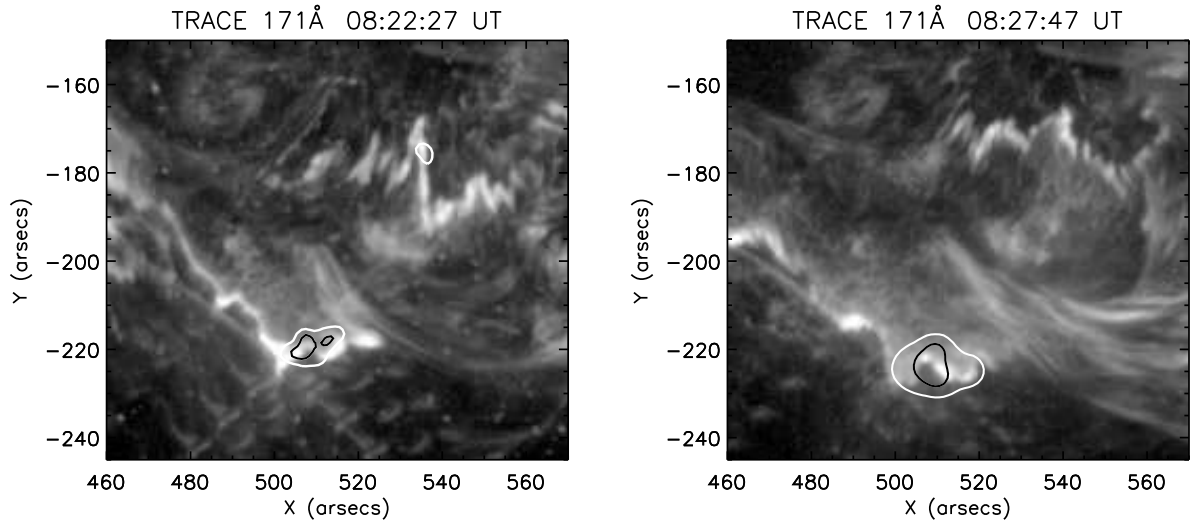}} 
\caption{TRACE 171~{\AA} filtergrams and overlay of RHESSI 20--60~keV images reconstructed with the Pixon algorithm
during two distinct HXR peaks associated with strong chromospheric evaporation flows. The RHESSI image integration times are 
08:22:48--08:23:36~UT (left) and 08:28:30--08:30:00~UT (right). The contours are 
at the 30\% (white) and 50\% (black) levels of each images's maximum.
Note that the RHESSI flare area used to determine the electron beam flux were derived within the 50\% contour of these images. 
 } \label{fig_rhessi_images}
\end{figure}

In order to obtain the energy flux density in fast electrons, we also need the cross-sectional area of the electron beam, 
which we estimate from the HXR sources. HXR footpoint sizes were determined from RHESSI images reconstructed over individual peaks in the 20--60~keV band 
using the Pixon and the MEM-NJIT algorithms. For the highest RHESSI peak during 08:22--08:24~UT the derived area is in the range
$ A \sim (2-4) \cdot 10^{17}$~cm$^2$, and for the small peak during around 08:29~UT, which is accompanied by the highest upflow velocities observed by CDS in the O\,{\sc v} line, we find $A \sim (4-8) \cdot 10^{17}$~cm$^2$. 
The uncertainties of about a factor of 2 were obtained from the different image reconstruction algorithms. In Fig.~\ref{fig_rhessi_images}
we show the two RHESSI Pixon images that we used for the source size determination.

With the energy flux in electrons above 20~keV obtained from the RHESSI spectral fits (see Fig.~\ref{fig_rhessi_params}), this gives for the highest RHESSI peak during 08:22--08:24~UT an energy flux density of $P_{20} = (4-7)\cdot 10^{10}$~erg~s~$^{-1}$~cm$^{-2}$, and for the peak at 
08:29~UT $P_{20} = (3-5)\cdot 10^9$~erg~s~$^{-1}$~cm$^{-2}$.

\section{Discussion}

The comparison of the energy flux densities in electrons that we derived from the RHESSI observations with the results of hydrodynamic simulations of 
the atmospheric response to electron beam heating \cite[e.g.,][]{1984ApJ...279..896N,fisher1985a,fisher1985b,fisher1985c,abbett1999,2005ApJ...630..573A} suggests that 
for the largest RHESSI peak during 08:22--08:24~UT the derived value of $P_{20} = (4-7)\cdot 10^{10}$~erg~s~$^{-1}$~cm$^{-2}$
is consistent with explosive chromospheric evaporation. \cite{fisher1985a} estimated from their simulations a threshold between 
gentle and explosive evaporation of $P_{20} \sim 10^{10}$~erg~s~$^{-1}$~cm$^{-2}$. The CDS measurements during this period show 
downflows in the range 20--40~km~s$^{-1}$ in the ``cool'' lines formed in the chromosphere and transition region, 
while we observe upflows with line-of-sight velocities up to about $-50$~km~s$^{-1}$ in the hot coronal Si~{\sc xii} line
(Fig.~\ref{fig_cds_short}). This flow behavior in the different layers of the flaring atmosphere matches qualitatively the expectations for explosive chromospheric evaporation. 
During the explosive evaporation process, strong pressure gradients build up in the chromosphere/lower transition region driving plasma flows
in both directions. Due to the larger mass and inertia of the chromosphere, the downflows are slower than the upflows of the hot plasma, establishing momentum
balance. 
The characteristics of the transition region lines that we observed during this highest peak is in line with the results of \cite{2005ApJ...625.1027K}, 
who detected short-lived O\,{\sc v} transition region downflows at the locations of H$\alpha$ flare kernels 
at the time of the strongest energy input (as evidenced by the peak of the GOES SXR flux derivative) in all four flares of their study. 
In addition, these authors report that the intermediate temperature between chromospheric evaporation upflows and downflows is close to Mg~{\sc ix} (1.0~MK). 

The relatively small upflow velocities observed in the Si~{\sc xii} line ($T\sim 2\times 10^6$~K) are basically in line 
with the recent hydrodynamic simulations by \cite{liu09}, who calculated the relation of 
the plasma flow velocities and the temperature. However, TRACE postflare loops (see last two panels in Fig.~\ref{fig_trace_rhessi}) 
indicate that actually the loop geometry relative to the CDS slit is not very favorable to measure line-of-sight velocities in the corona, 
since the coronal loops cross the slit with a strong transversal component, and the coronal velocities derived from the Si~{\sc xii} line 
may be underestimated. We also note that the intensity behavior in the Si\,{\sc xii} line (impulsive emission spike followed by a gradual increase, i.e.\ a kind of 
Neupert-effect-type plasma response; see Fig.~\ref{fig_cds_short}, bottom panels) indicates that we observe the emission of 
plasma accumulating in the corona along the line-of-sight in the selected CDS pixels, due to the chromospheric evaporation process. 
This is quite different to the impulsive peaks observed in He\,{\sc i} and in O\,{\sc v} (Fig.~\ref{fig_cds_short}, middle panels), 
where we observe the emission from narrow layers
in the chromosphere and transition region, and the intensity increase occurs instantaneously and correlated with the energy input sequence.

More interesting because difficult to reconcile with the ``simple'' explosive versus gentle evaporation picture is the 
plasma behavior when compared to the electron beam parameters derived for the small but distinct HXR burst at 08:29~UT. For the 
energy flux density in electrons we find $P_{20} = (3-5)\cdot 10^9$~erg~s~$^{-1}$~cm$^{-2}$, which is a heating flux too small to drive 
explosive chromospheric evaporation according to the model results of \cite{fisher1985a}. This would suggest that gentle chromospheric evaporation is at work,
and the transition region is expected to slowly ablate with velocities of the order of some ten km~s$^{-1}$.
However, at that period we observe the highest upflow velocity in the O{\,\sc v} transition region line during the event, with line-of-sight velocities 
up to $-280$~km~s$^{-1}$. 
We note that we cannot exclude the possibility that the power in the electron beam is actually higher than we derived in our analysis, 
since this quantity critically depends on the cross-sectional area of the HXR footpoints and the low-energy cut-off to the accelerated electron spectrum, 
which are both difficult to be exactly derived from the observations. The value we derived for the energy flux in electrons 
is actually close to the threshold between gentle and explosive evaporation. However, even in case that the energy flux 
in electrons would indeed suffice to drive explosive chromospheric evaporation, 
the CDS observations cannot straightforwardly be reconciled
with the hydrodynamic modeling results of the flaring atmosphere in terms of explosive versus gentle evaporation 
because the velocity of 280~km~s$^{-1}$ in the O\,{\sc v} transition region line is directed upward. 
It is also considerably higher than the upper limit derived by 
\cite{fisher1985b} for the expansion velocity of explosive chromospheric evaporation, which is about 2.35 times the sound speed of the evaporating plasma. 
The formation temperature of the O{\,\sc v} spectral line is $2.7 \times 10^5$~K and the corresponding sound speed $\sim$70~km~s$^{-1}$, 
giving a limit to the expansion velocity of about 140~km~s$^{-1}$.

We also stress that during these times, we observe in the O\,{\sc v} transition region line contributions from both a stationary and a high-velocity 
upflowing plasma volume, in single CDS pixels which have a size as small as $2'' \times 1.7''$, corresponding to an area of 
$1.8 \times 10^{16}$~cm$^{2}$. This implies intriguing dynamics on fine-structured scales of the flaring atmosphere.
Double component spectra in transition region lines during a flare have been observed before 
in individual pixels during high-cadence CDS sit-and-stare spectroscopic observing mode \citep{brosius2003} but --- at least to our knowledge --
have not been reported and studied in detail
in spectra rastering over the flare region. This is most probably due to the very dynamic nature of the transition region during solar flares, 
which is difficult to capture in raster mode. It is also remarkable that this distinct intensity peak showing high upflow velocities in the O\,{\sc v} 
transition region line is not accompanied by an intensity peak or plasma flows in the chromospheric He\,{\sc i} line. 
This suggests that most of the electron beam energy is deposited in the transition region layers causing the strong plasma upflows there, and does not 
reach the chromosphere.

Finally, we note that the light curves extracted in individual CDS pixels in the He\,{\sc i} and O\,{\sc v} spectral lines show several distinct peaks during the 
impulsive flare evolution. 
This implies that either a) the same area of the solar atmosphere is energized several times during the event (which is in conflict with 
the standard eruptive flare model, where magnetic reconnection driven by the erupting CME activates in sequence different magnetic loops further and further away from the magnetic inversion line), or b) different loops are energized within the same pixel (multi-thread scenario) but cannot be distinguished within the spatial resolution of the instruments. The multi-thread interpretation could also explain the double component spectra observed in individual CDS pixels.
In addition, in the multi-thread scenario the electron beam flux on individual threads would be larger than we estimated, and could potentially explain the large range of 
upflow velocities that is observed with each hard X-ray burst.

\section{Conclusions}
We presented high-cadence CDS spectroscopy combined with high-cadence imaging of the solar atmosphere in comparison with the energy input by fast electrons as derived from RHESSI HXRs in an M2.5 flare. For the flare impulsive phase, the plasma flow behavior (simultaneous downflows in the ``cool'' He\,{\sc i} and O\,{\sc v} lines 
formed in the chromosphere and transition region, and upflows in the hot coronal Si~{\sc xii} line) as well as the derived energy deposition rate by electron beams are consistent with explosive chromospheric evaporation. 
However, for a later distinct HXR burst, where the strongest energy deposition site is exactly located on the CDS slit, the situation is much more complex. 
The energy input by electrons is about an order of magnitude smaller than during the flare peak and too small to drive explosive chromospheric evaporation. 
However, at the same time we observe the highest upflow velocities in the O\,{\sc v} transition region line during the event, up to 280~km~s$^{-1}$, in addition
to the contribution of a stationary plasma component within the same CDS pixel. The upflow velocities are much too high for gentle ablation of the transition region. 
These findings indicate that the flaring transition region is extremely dynamic, complex and fine-structured on scales that are smaller than can be resolved with present instrumentation, and is thus probably not adequately captured by single-loop hydrodynamic simulations of the flaring atmosphere.

\acknowledgments
A.M.V. gratefully acknowledges support of the Austrian Science Fund (FWF): P20867-N16. M.T. is a recipient of an APART-fellowship of the Austrian Academy of Sciences at the Institute of Physics, University of Graz (APART 11262). This work was supported by the scientific exchange program SK-AT-0004-08 of
the ``\"Osterreichischer Austauschdienst" \"OAD (Austria) and the Slovak Research and Development Agency SRDA (Slovakia) as well as by the SRDA project
APVV-0066-06. Data were acquired in the frame of the SOHO Joint Observing Program JOP171 operated in cooperation with the TRACE satellite, the SOHO satellite and the ground-based 
DOT, Hvar and Kanzelh\"ohe telescopes. The authors are grateful for the cooperation of the SOHO, TRACE, and DOT teams during the run of JOP171 in July 2006.
SOHO is a project of international cooperation between ESA and NASA. TRACE is a mission of the Stanford-Lockheed Institute for Space Research and part 
of the NASA Small Explorer Mission. RHESSI is a NASA small explorer mission. A.M.V. thanks Luca Teriaca, Ryan Milligan, Wei Liu and Ken Phillips 
for interesting and insightful discussions. We are also grateful to the anonymous referee for his/her insightful comments which helped to improve the paper.

\bibliographystyle{apj}

\end{document}